\documentstyle[sprocl,epsf]{article}
\input psfig
\psfull
\begin{document} 
\null
\vspace*{-24pt}
\centerline
{DPF{\,}'96: The Minneapolis Meeting (11-15 August 1996) pages 868--870.}
\bigskip
\title{$D-\pi$ Production Correlations 
in 500 GeV $\pi^--N$ Interactions}
\author{ K. GOUNDER and L. CREMALDI }
\author{ for the Fermilab E791 Collaboration}
\address{Dept.\ of Physics and Astronomy, 
University of Mississippi,
Oxford, MS 38677}

\maketitle\abstracts{ 
We study the charge correlations between charm mesons 
produced in  500 GeV $\pi^--N$ interactions and the charged pions
produced closest to them in phase space. With 110,000 fully 
reconstructed {\em D} mesons from experiment E791 at Fermilab, the 
correlations are studied as functions of the  $ D\pi - D $ mass difference 
and of Feynman $ x $. We observe significant correlations which appear to 
originate from a combination of sources including fragmentation
dynamics, resonant decays, and charge of 
the beam. }

In the production of a $D$ meson the charge of pions formed nearby in 
phase space may be highly correlated with the charge of the $D$ meson. 
For example, this $associated~pion$ can be produced in the fragmentation
dynamics\,\cite{the:ros2} of a $ c $ quark combining with a $ \overline d $ 
from a 
$ d -\overline d$ vacuum pair, forming a $ D^+ $, and the remaining $ d $
forming an associated pion nearby in phase space.
Thus, $ D^+( D^{*+}) \pi^- $, $D^-(D^{*-}) \pi^+$, 
and $ D^0 \pi^+ $ $ \overline {D {}^0} \pi^- $ would be favored over 
combinations with opposite pion charges.
Resonant decays\,\cite{the:ros1} and the charge of the beam particle are also 
sources of charge correlations. Using a $ \pi^- $ beam can lead to the 
association of both charm mesons and anticharm mesons with negative pions, 
especially in the forward (beam) direction.
By comparing the charge correlations of different species of charm mesons
and antimesons with associated pions, and by studying them as functions
of Feynman $ x $ ($x_F$), one can hope to disentangle some of these processes.

We use well identified $D^0 \to K^- \pi^+$, $D^+ \to K^- \pi^+ \pi^+$, 
and $D^{*+} \to D^0 \pi^+$ signals (and their charge conjugate decays) from 
experiment E791 at Fermilab for this study. A silicon vertex detector and
two {\v C}erenkov counters are used to locate decay vertices and identify
pions and kaons for this analysis. Details of the spectrometer and event 
selection are found elsewhere.\,\cite{ref:e791} 
The final $D$ meson sample sizes used for this analysis are
$22587\pm210$, $24237\pm216$, $24569\pm204$, $29649\pm238$, 
$4997\pm84$, and $6048\pm93$ events, for 
$D^0$, $\overline{D {}^0}$, $D^+$, $D^-$, $D^{*+}$, and $D^{*-}$, 
respectively. 

For each {\em D} found in an event, all identified pion tracks originating 
from the primary vertex  are combined with the {\em D}. Among these combinations, the pion 
that forms the smallest invariant mass difference ($\Delta m_{min}$) with the {\em D} 
is selected as the candidate associated pion. 

We define the correlation parameter $\alpha$ as
\begin{equation}
 \alpha(D) \equiv \frac{\sum N_i(D\pi^q)-\sum N_i(D\pi^{-q})}
{\sum N_i(D\pi^q)+\sum N_i(D\pi^{-q})}
 \label{eq:aeq1}
\end{equation}
\noindent where $ q $ =  +, $ - $, $ - $, +, $ - $, + for $D = D^0$, 
$\overline {D {}^0}$ , $ D^+$, $ D^-$,
$D^{*+}$, and $D^{*-}$, respectively, and $\sum N_i(D\pi^q)$ 
denotes the number of charm mesons for which the selected pion has the charge
$ q $. 

The correlation parameters for 
background-subtracted signals and background regions are listed in Table 1. 
The signal correlations differ significantly from the background correlations. 
These raw correlations must be corrected for the occurrence of falsely
reconstructed tracks (ghosts), momentum smearing in bins of $\Delta m_{min}$,
detector acceptance, and dilution of the correlation by wrong track
selection. We use a Monte Carlo simulation of the experiment and
the LUND event generator (PYTHIA 5.7/JETSET 7.3)\,\cite{ref:lmc} 
to model these effects and extract the final results (Table 1), within a matrix formalism
explained elsewhere.\cite{ref:prl} A fundamentally different model of hadron 
production could change the corrections; an effect not accounted for 
in Table 1. 

\begin{table}[btp]
\caption{
The $ x_F $- and $\Delta m $-integrated correlation parameters $\alpha$ defined
in Eq.~(1) for the background-subtracted signals prior to correction,
for the corresponding backgrounds, and for the signals after correction
using the matrix correction technique.
}  
\label{tab1}
\renewcommand{\arraystretch}{1.2}
\vskip 2pt
\begin{tabular}{lccc} \hline
Charm & Signal $ \alpha $  & Background $ \alpha $
      & Corrected Signal $ \alpha $ \\
\hline
$D^0$ & $0.13\pm0.01$ & $-0.04\pm0.01$ & $0.12\pm0.03\pm0.04$ \\
${\overline {D {}^0}}$ & $0.18\pm0.01$ & $0.04\pm0.01$ &
$0.42\pm0.02\pm0.03$ \\
$D^+$ & $0.18\pm0.01$ & $0.10\pm0.01$ & $0.45\pm0.03\pm0.03$ \\
$D^-$ & $0.08\pm0.01$ & $0.02\pm0.01$ & $0.03\pm0.03\pm0.04$ \\
$D^{*+}$ & $0.15\pm0.02$ & $0.08\pm0.03$ & $0.33\pm0.05\pm0.03$ \\
$D^{*-}$ & $0.08\pm0.02$ & $0.02\pm0.03$ & $0.15\pm0.05\pm0.04$ \\
\hline
\end{tabular}
\end{table}

To investigate beam-related effects in more detail, we study the 
$x_F$ dependence of the $D^{+}$ and $D^{*+}$ correlations. 
In Fig. 1, we plot $ \alpha $ as a function of $x_F$, 
for both particle and antiparticle for $D^+$ and $D^{*+}$. 
\noindent
We observe that $\alpha(D^+)$ rises slightly with $x_F$ but $\alpha(D^-)$ 
falls sharply to negative values for $x_F > 0.2$.
In both cases, the $ D $'s are more likely to be associated
with $ \pi^- $'s at high $ x_F $ where beam effects seem to be important.
There appears to be less dependence of $\alpha$ on $x_F$ for the $D^{*\pm}$.

Detailed Monte Carlo studies 
suggest minimal or no beam-related effects when the combined particle and 
antiparticle correlations are computed. We calculate  the 
combined correlation parameters to be
$\alpha(D^0,\overline {D {}^0}) = 0.29 \pm 0.02 \pm 0.03$,
$\alpha(D^+,D^-) = 0.21 \pm 0.02 \pm 0.03$, and
$\alpha(D^{*+},D^{*-}) = 0.23 \pm 0.04 \pm 0.03$.  
These results indicate that fragmentation dynamics and resonant decays 
produce substantial correlations between  $ D $ mesons and their 
associated pions. All three combined correlation levels are approximately 
equal. 

In summary, we observe significant production correlations between
$ D $ mesons and their associated pions. Beam-related and other  
fragmentation effects play an important role. Further studies are 
needed to unravel the contributions.  

This work was supported in part by U.S.\ DOE DE--FG05--91ER40622.

\begin{figure}[htbp]
\epsfxsize=2.0in
\hbox{\hfill\hskip+0.75in\epsffile{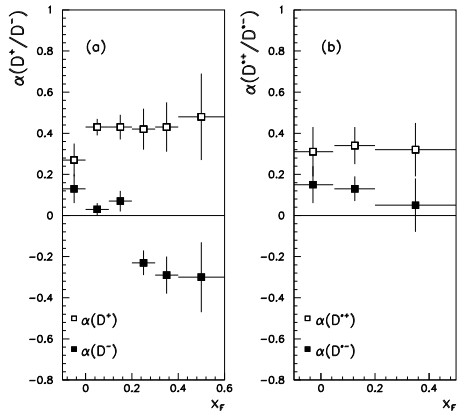}\hfill}
\caption{The corrected correlation parameter $\alpha$ as a function of
$x_F$ for (a)$D^+$ and (b)$D^{*+}$.  The parameter $\alpha$ is defined
in Eq. (1) in the text. The error bars correspond to the statistical and 
systematic uncertainties added in quadrature.} 
\label{fig:xf}
\end{figure}

\end{document}